\documentclass[twocolumn,floatfix,aps,nofootinbib]{revtex4}
\usepackage{epsfig}

\def\lp {\left( }
\def\rp {\right) }
\def\lb {\left[ }
\def\rb {\right] }
\def\lc {\left\{ }
\def\rc {\right\} }

\def\nn {\nonumber}

\def\beq{\begin{equation}}
\def\eeq{\end{equation}}
\def\bea{\begin{eqnarray}}
\def\eea{\end{eqnarray}}
\def\ni{\noindent}

\def\ub {\bar u}

\def\Pb {\bar{\Pi}}

\def\mb {M}

\def\sb {\bar{s}}

\def\g{\gamma}

\def\L {\Lambda}
\def\m{\mu}
\def\n{\nu}

\def\p{\pi}
\def\P{\Pi}

\def\s{\sigma}
\def\S{\Sigma}

\def\bp {\mbox{\boldmath $p$}}

\begin{document}

\title{$\pi \Lambda$ phase shifts and $CP$ Violation in  
${ \Xi\rightarrow \pi \Lambda}$ Decay}

\author{C.C. Barros Jr.}

\affiliation{Instituto de F\'{\i}sica, Universidade de S\~{a}o Paulo,\\
C.P. 66318, 05315-970, S\~ao Paulo, SP, Brazil}

\date{\today}

\begin{abstract}
The $CP$ violation asymmetry parameters in $\Xi \rightarrow\pi\Lambda$ 
nonleptonic decays
are presently being measured by the HyperCP experiment.
In the study of these $CP$ violation parameters,  the strong $S$ and $P$ 
phase shifts for the $\pi\Lambda$
final-state interactions are needed.
In this work,
these phases are calculated using an effective chiral
Lagrangian model, that 
considers 
 $\Sigma$, $\Sigma^*$(1385), and the $\sigma$-term, in the 
intermediate states. The $\sigma$-term is calculated in terms of the scalar 
form factor of the baryon.
\end{abstract}

\maketitle

\vspace{5mm}

\section{introduction}

The $CP$ violation phenomenon is an important aspect in the understanding of 
the nature. Differences in the behavior of particles and antiparticles provide 
 informations that may clarify many aspects of the current particle
physics,
 as for example, if
 $CP$ violation appears as a consequence of the standard model, by
 the  Kobayashi-Maskawa mechanism \cite{KM}, 
or if this violation appears from  
physics  beyond the standard model,  such as the superweak model  \cite{sw} or 
others \cite{WH}, \cite{pen}.

Even in the cosmology, when the baryogenesis is studied, and the 
matter-antimatter asymmetry in the universe
is investigated, the ratio 
$n_B/n_\gamma=(5.5\pm 0.5)\times10^{-10}$ 
between the baryon number density, $n_B$, and the photon number
density, $n_\gamma$,
 is essentially a $CP$ violation  observable
and is incompatible with the  Kobayashi-Maskawa model \cite{nir}.

Experimentally, the breakdown of $CP$ invariance
 has been observed in two kinds of weak $K$ decays, and recently in weak 
$B$ decays also, \cite{nir}-\cite{aba}. In the three observed cases, the
results are in accord with the Kobayashi-Maskawa picture.
 Another system where $CP$ violation is 
expected to occur, is the nonleptonic decay of hyperons \cite{OK}, \cite{PA}. 
This effect, in the decay sequence $\Xi^-\rightarrow 
\Lambda \pi^-$,
 $\Lambda \rightarrow N\pi$ and in the respective ${\overline{\Xi}}^+$ one,
has been measured in the E756 experiment, 
that obtained $A_{\Xi\L}\simeq A_\Xi+A_\L=0.012\pm 0.014$ 
 \cite{luk}, \cite{luk2}  and  
$S-P=3.17^o\pm 5.28^o\pm 0.73^o$  \cite{E756}, where $S$ and $P$ are the
strong phases at the $\Xi^-$ mass.
 These results are still inconclusive, concerning
to prove the $CP$, violation, but the same decay 
is presently being measured by the 
 HyperCP experiment, that is also dedicated to  look for CP violation in
$\Xi^-$ decay, but with high sensitivity.

Some authors studied this decay \cite{don}-\cite{meis} and in order to 
determine 
the asymmetry decay parameters, the phases $\delta_l$,
$\phi_l$ in the final-state $\pi$-barion
 interactions  are needed. The weak $\phi_l$ phases, are the 
responsible for the $CP$ violation and have been calculated,
 in \cite{ta2}, in the framework of the standard model, and in 
\cite{xh1}, \cite{xh2}, \cite{ta11} where new physics was considered.
The $\Omega^-\rightarrow \pi\Xi$ decays have also been studied \cite{CW}, 
\cite{ta3}, and the resultant strong phase differences, now in the 
final $\pi\Xi$ interactions,  were 
significantly bigger.

The strong  $\pi N$ phases,  in the $\Lambda\rightarrow\pi N$ and 
$\Sigma\rightarrow\pi N$  decays are very well known \cite{PS}, and 
can be described by chiral models, \cite{LG}, \cite{PiN}.
So, it seems reasonable  to calculate the phases in the other hyperon decays 
($\Xi$ and $\Omega$), using chiral models also.
The strong phase difference $\delta_P - \delta_S$ in the $\pi\Lambda$
interaction
 has been estimated 
theoretically in some works \cite{Kam}-\cite{meis}, \cite{keis}, 
\cite{BH}.

The recent results from the E756 collaboration, and the future results
from the HyperCP experiment, motivates theoretical studies in the
field.
In this paper, the strong phases $S$ and $P$, in the 
$\pi\Lambda$ interaction will be estimated, with a model that was used 
to calculate the strong $\pi\Xi$ phase shifts in the $\Omega^-$ decay 
\cite{CW}.  This model is an improvement of the one proposed in
\cite{BH}, to study the strong $\pi Y$ interactions, and 
 is based on chiral Lagrangians, with $\S$ and $\S^*$(1385) in
the intermediate states. The inclusion of the 
scalar form factors $\sigma(t)$
of the hyperons  will be made, in a way similar to the one found in
\cite{MAN}, \cite{CM}, \cite{CW}. 
Another improvement that will be made, is the utilization of 
 the  $\Lambda\pi\Sigma$ coupling constants, obtained in 
\cite{LOI} from the study of hyperonic atoms.

This paper will show the following contents: In section II, a brief
review of the asymmetry parameters of the
$\Xi^-$ decay will be made. In section III the strong phases  
 in the $\pi\Lambda$ interactions will be calculated. The results of
 this work, and the ones of the other models available will be shown 
in section IV.

\section{Nonleptonic ${\bf \Xi^-}$ Decay}

In the  $\Xi^- \rightarrow\Lambda\pi^-$ decay, the amplitudes with $L$=0
($S$) and $L$=1 ($P$) may be parametrized as 
\beq
S=|S|e^{i(\delta_S+\phi_S)}
\label{2.12}
\eeq
\beq
P=|P|e^{i(\delta_P+\phi_P)}   \   \   ,
\label{2.13}
\eeq

The  observables of interest in the study of $CP$ violation are the asymmetry 
parameters $\alpha$, $\beta$ and $\gamma$ , which  can be expressed in terms
 of the $S$ and $P$ amplitudes,
\beq
\alpha=2\ {\rm Re}(S^*P)/(|S|^2+|P|^2)
\eeq
\beq
\beta=2\ {\rm Im}(S^*P)/(|S|^2+|P|^2)
\eeq
\beq
\gamma=(|S|^2-|P|^2)/(|S|^2+|P|^2)
\eeq

\noindent
and obeys the relation
\beq
\alpha^2+\beta^2+\gamma^2=1  \  \  .
\eeq

In the systems where $CP$ conservation exist, the $CP$ asymmetry parameters
\beq
A={\alpha+{\overline \alpha}\over \alpha-{\overline \alpha}}
\eeq

\ni
and
\beq
B={\beta+{\overline \beta}\over \beta-{\overline \beta}}
\eeq

\ni 
vanish, since $\alpha$=-${\overline \alpha}$ and $\beta$=-${\overline \beta}$. 

These asymmetry parameters are approximately
\beq
A(\Xi_-^-)=A_\Xi =
-{\rm tan}(P-S)
{\rm tan}(\phi_P^\Lambda-\phi_S^\Lambda)  \  \  ,
\label{1.30}
\eeq

\ni
and
\beq
B(\Xi_-^-)= B_\Xi = 
{\rm cot}(P-S)
{\rm tan}(\phi_P^\Lambda-\phi_S^\Lambda)  \  \  .
\label{1.31}
\eeq

\noindent
In the $\Lambda\rightarrow\p N$, the $A_\L$ and $B_\L$ may be defined
in a similar way.

In the next section we will calculate the phase shifts $\delta_l$, that
are needed to estimate $A$ and $B$.

\section{Low energy ${\pi\Lambda}$ interaction}

In this section, 
the low energy $\pi\Lambda$ interaction will be studied with the model 
proposed in \cite{CW}. 
 In this model, the interactions are considered to be described by 
effective chiral Lagrangians, that in the  $\pi\Lambda$ interactions are

\begin{eqnarray}
{\cal{L}}_{\Lambda\pi\Sigma} &=& {g_{\Lambda\pi\Sigma}\over 2m_\Lambda}\lc
 {\overline\Sigma}
\gamma_\mu \gamma _{5} \vec \tau\Lambda \rc .\partial^\mu \vec \phi  \\
{\cal{L}}_{\Lambda\pi\Sigma^*} &=& g_{\Lambda\pi\Sigma^*} \lc 
{{\overline{\Sigma}^*}^\mu}
\lb g_{\mu\nu} - (Z+{1\over 2})\gamma_\mu \gamma _\nu  
\rb \vec\tau\Lambda \rc .\partial ^\nu \vec \phi \nn \\
&&
 + H.c.  \  \ , 
  \label{3.1} 
\end{eqnarray}

\noindent
where $\Lambda$, $\Sigma$, $\Sigma^*$ and $\vec\phi$ are the lambda, the
sigma, 
the resonance 
$\Sigma^*$(1385), and the pion fields. $Z$ is the off-shell parameter 
\cite{PiN}.

This model is supported by the fact that the low energy $\pi N$ interactions 
are quite well described by similar Lagrangians \cite{PiN}.
In analogy with the $\pi N$ interactions, where the $\Delta$(1232) resonance 
dominates the amplitudes, the intermediate 
$\Sigma^*$(1385) are also included. Fig. 1 shows  the diagrams to be 
considered. 
The inclusion of the $\sigma$ term (Diagram 1c) is needed by the fact that in 
the  $\pi N$ interaction, if one make the calculations  only with the 
tree diagrams, with $N$ and $\Delta$ in the intermediate states, 
the accord
with the experimental data is not so good, but the
the inclusion of the $\sigma$ exchange \cite{PiN}, improves this accord.

The spin 3/2 propagator for a mass $M$ particle, is then 
\bea
G^{\m\n}(p)  &=& - \; \frac{(\not\!\!{p}+M)}{p^2-M^2}\lp g^{\m\n}
 -\frac{\g^\m\g^\n}{3}
 \right. \nn \\
& &\left.
-\frac{\g^\m p^\n}{3M} + \frac{p^\m \g^\n}{3M} - \frac{2 p^\m p^\n}{3M^2}\rp \;.
\label{2.3}
\eea

\begin{figure}[hbtp]
\centerline{
\epsfxsize=80.mm
\epsffile{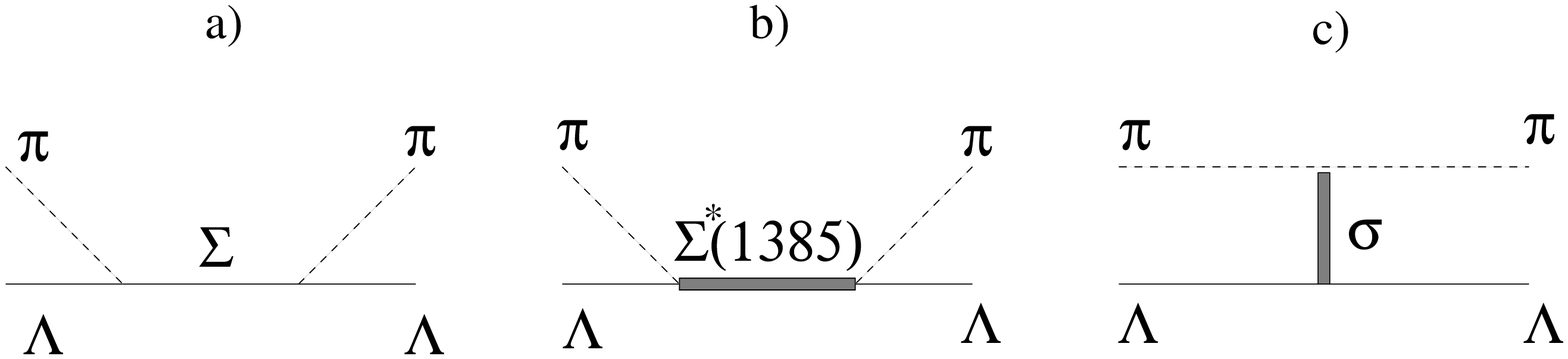}}
\caption{Diagrams to $\pi\Lambda$ Interaction.}
\end{figure}


The $\pi\Lambda$ scattering matrix will have the general form
\beq
T_{\pi\Lambda}^{ba}= \overline{u}(\vec{p}\prime ) \lc A + 
{(\not\!k + \not\!k')\over 2}B\rc  
 u(\vec p)      \    \     ,
\eeq

\ni
where $k$ and $k'$ are the initial and final $\pi$ momenta. 
The contributions from Fig. 
1(a) (intermediate $\Sigma$) are
\begin{eqnarray}
A_\Sigma &=& {g_{\Lambda\pi\Sigma}^2(m_\Lambda+m_\Sigma)}\lc {\frac{%
s-m_\Lambda^2}{s-m_\Sigma^2}}+{\frac{u-m_\Lambda^2} {u-m_\Sigma^2}}\rc\ ,
 \nonumber \\
B_\Sigma &=& {g_{\Lambda\pi\Sigma}^2}\lc {\frac{m_\Lambda^2-s-2m_%
\Lambda(m_\Lambda+m_\Sigma)}{s-m_\Sigma^2}}  \nonumber \right. \\
&&
\left.  \  \  \  \ \ \ \   \  \
+{\frac{2m_\Lambda(m_\Lambda+m_\Sigma)+ u-m_\Lambda^2}{u-m_\Sigma^2}}
\rc \ \ .
\end{eqnarray}

\noindent
The contribution from Fig. 1(b), the interaction with the intermediate
resonance $\Sigma^*$, is
\begin{eqnarray}
A_{\Sigma^*} &=& {\frac{g_{\Lambda\pi\Sigma^*}^2}{3m_\Lambda}}\lc 
{\frac{\nu_r}{\nu_r^2-\nu^2}}\hat A-{\frac{m_\Lambda^2+m_\Lambda m_{\Sigma^*}}{m_{\Sigma^*}^2}} \right.  \nonumber \\ 
&&
\left.
\hspace{1.cm}\times\lp2 m_{\Sigma^*}^2+m_\Lambda m_{\Sigma^*}
-m_\Lambda^2+2\m^2\rp \right. \nonumber \\
&&
\left.
+{\frac{4m_\Lambda}{m_{\Sigma^*}^2}}\lb (m_\Lambda\!+
m_{\Sigma^*})Z\!+\!(2m_{\Sigma^*}\!+m_\Lambda)Z^2\rb k.k^{\prime}
\rc\ , \nonumber \\
B_{\Sigma^*} &=& {\frac{g_{\Lambda\pi\Sigma^*}^2}{3m_\Lambda}} 
\lc{\frac{\nu}{\nu_r^2-\nu^2}}\hat B - {\frac{8m_\Lambda^2\nu Z^2} 
{m_{\Sigma^*}^2}}\rc\ ,
\end{eqnarray}

\ni
where $\nu$ and $\nu_r$ are
\begin{eqnarray}
\nu &=& {\frac{s-u}{4m_\Lambda}} \\
\nu_r &=& {\frac{m_{\Sigma^*}^2-m_\Lambda^2-k.k^{\prime}}{2m_\Lambda}} \ \ ,
\end{eqnarray}

\ni
 and 
\begin{eqnarray}
\hat A&=&{\frac{(m_{\Sigma^*}+m_\Lambda)^2-\m^2}{2m_{\Sigma^*}^2}}
\lc 2m_{\Sigma^*}^3-2m_\Lambda^3-2m_\Lambda m_{\Sigma^*}^2
\right.  \nonumber \\
&&
\left.
-2m_\Lambda^2m_{\Sigma^*}\!\!+\!\m^2(2m_\Lambda\!\!-\!m_{\Sigma^*})
\rc+{\frac{3}{2}}(m_\Lambda\!\!+\!m_{\Sigma^*})t\ ,  \nonumber \\
\hat B &=& {1\over 2m_{\Sigma^*}^2}\lbrack (m_{\Sigma^*}^2-
m_\Lambda^2)^2
-2m_\Lambda m_{\Sigma^*}(m_{\Sigma^*}+m_\Lambda)^2 \nonumber \\
&&
+6\m^2m_\Lambda (m_{\Sigma^*}+m_\Lambda)
-2\mu^2(m_{\Sigma^*}+m_\Lambda)^2+\mu^4\rbrack + {3\over 2}t \ ,  \nonumber \\
&& \label{3.2} 
\end{eqnarray}

\ni
 where $\mu$ is the pion mass.

In \cite{PiN}, \cite{BH} the $\sigma$ term (diagram 1.c) was
   parametrized as
\begin{eqnarray}
A_\sigma &=& a+bt   \nonumber  \\
B_\sigma &=& 0 \   \   ,
\end{eqnarray}

\ni
with the $a$ and $b$ parameters determined by the experimental
$\p N$ data. In \cite{MAN}, \cite{CM}, this contribution was calculated, relating
 it with the scalar form factor of the baryon.  
In fact, the $\sigma$ term represents the exchange of a scalar isoscalar 
system in the $t$-channel, and at large distances is dominated by triangle 
diagrams (Figure 2)
involving the exchange of 2 pions  \cite{LW}.

\begin{figure}[hbtp]
\centerline{
\epsfxsize=90.mm
\epsffile{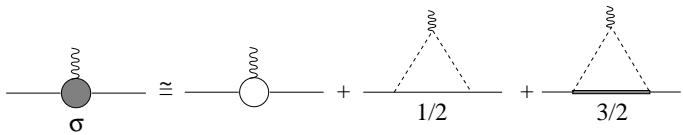}}
\caption{ The scalar form factor (gray blob)  receives contributions from tree
 interactions (white blob)
and triangle diagrams with spin $1/2$ and $3/2$ intermediate states.}
\end{figure}

The scalar form factor for a spin $1/2$ baryon $B$ is defined 
in terms of the chiral symmetry breaking Lagrangian  ${\cal L}_{sb}$ by
the expression
$ <B(p') | - {\cal L}_{sb} | B(p) > \equiv \s(t) \; \ub(\bp') \;u(\bp) $,
and as it can be seen in Fig. 2, it will receive contributions 
from spin 1/2 ($\Sigma$) and spin 3/2 ($\Sigma^*$) intermediate states. These 
contributions can be expressed in terms  of the loop integrals $\Pi$
defined
in Appendix B
(see \cite{CM}, \cite{MAN} for details on the calculations)

\bea
&& \s_{1/2}(t) = \frac{\m^2}{(4\p)^2}\lp \frac{g_{\Lambda\p\Sigma}}
{2m_\Lambda}\rp^2
  (m_\Lambda\!+\!m_{\Sigma^*})
\lb \P_{cc}^{(000)}- \right. \nn\\
&&\left.
\frac{m_\Lambda^2\!-\!m_{\Sigma^*}^2}{2m_\Lambda\m}\;\P_{\sb c}^{(000)} -\;
\frac{m_\Lambda+m_{\Sigma^*}}{2m_\Lambda}\;\P_{\sb c}^{(001)} \rb\;,
\label{2.8}\\[4mm]
&& \s_{3/2}(t) = \frac{\m^2}{(4\p)^2} \lp 
\frac{g_{\L\p\S^*}}{2m_\L}\rp^2 \frac{1}{6m_{\S^*}^2}
\lc - \lb (m_\L\!+\!m_{\S^*})^2 \right.\right. \nn\\
&&\left.\left.
\times(2m_{\S^*}\!-\!m_\L) + 2 \m^2 (m_\L\!+\!m_{\S^*})
+ (\m^2\!-\!t/2) \; m_\L  \rb \P_{cc}^{(000)}\right.
\nn\\[2mm]
&& \left.
- 2 \m^2 m_\L  \; \Pb_{cc}^{(000)}
+ \lb (m_\L^2\!-\!m_{\S^*}^2) (m_\L\!+\!m_{\S^*})^2 
(2m_{\S^*}\!-\!m_\L)\right.\right.
\nn\\[2mm]
&& \left.\left. 
+ 2\m^2 (m_\L\!+\!m_{\S^*})(m_\L^2\!-\!m_{\S^*}^2) \right.\right.
\nn\\[2mm]
&& \left.\left.
+ 6  (\m^2\!-\! t/2) m_{\S^*}^2 (m_\L\!+\!m_{\S^*})
 -\m^4 (2m_{\S^*}\!+\!m_\L) \rb \frac{\P_{\sb c}^{(000)}}{2m_\L\m}
 \right.
\nn\\[2mm]
&& \left. 
+ \lb (m_\L\!+\!m_{\S^*})^2 (4m_\L m_{\S^*}\!-\!m_\L^2\!-\!m_{\S^*}^2) 
+ 6 m_{\S^*}^2 (\m^2\!-\! t/2) 
\right.\right.
\nn\\[2mm]
&& \left.\left.
- 2 \m^2 (m_\L\!+\!m_{\S^*}) (2m_{\S^*}\!-\!m_\L) 
- \m^4 \rb
\frac{\P_{\sb c}^{(001)}}{2m_\L} \rc\;.
\label{2.9}
\eea

\ni
The total amplitude is then
\beq
T=T_\L+T_{\S^*}+T_\s
\eeq

\ni
If one calculates the amplitudes in the $\p\L$ center-of-mass frame, defining
the momentum, $\kappa$, the scattering angle,  $\theta$ and $x=\cos \
\theta$, the partial wave amplitudes $a_{l\pm}$ can be obtained with
equation (\ref{apar}). The phase shifts are calculated by
unitarization
of the amplitudes  \cite{EB}  with
\beq
\delta_{l\pm} = {\rm tg}^{-1}(\kappa\ a_{l\pm})  \  \  .
\eeq

\noindent
The parameters used can be found in 
 in \cite{PDG} and are
 $m_\Lambda$=1.115 GeV, $m_{\Sigma^*}$=1.384 GeV, 
 The coupling constant $g_{\Lambda\pi\Sigma}=12.92\pm 1.2$, is a recent value, 
obtained in  \cite{LOI}, and
the coupling constant $g_{\Lambda\pi\Sigma^*}$ 
 was calculated in \cite{BH}, and the value obtained was
 9.38 ${\rm GeV}^{-1}$. We will also use the standard value $Z$=-0.5.

The numerical results of the  phase shifts at $\sqrt{s}=m_{\Xi^-}$ are 
\beq
 -1.30^o\leq S\leq -1.11^o
\eeq  

\ni and
\beq
-1.83^o\leq P\leq -1.81^o   \  \   .
\label{3.20}
\eeq

\section{Summary and conclusions}

In the previous sections
the strong $S$ and $P$ phase shifts for the $\Xi^-$ decay
have been calculated
at the $\Xi^-$ mass   with a model based on chiral effective Lagrangians,
including the contributions from the diagrams of Fig. 1, and the diagram
1c is calculated with the diagrams of Fig. 2, that is essentially the
same model used in \cite{CW} to estimate the strong $\p\Xi$ phases.
The numerical values of the phases are  $-1.30^o\leq S\leq -1.11^o$ and
$-1.83^o\leq P \leq -1.81^o$ that gives $-0.71^o\leq P-S\leq -0.51^o$, that 
are smaller in magnitude
then the values obtained in \cite{BH},   $S=-4.69^o$, $P=-0.36^o$
and $P-S=4.3^o$ where, 
the $\s$ term was just considered as a parametrization. If compared
with the experimental results from \cite{E756} 
$P-S=3.17^o\pm 5.28^o\pm 0.73^o$, both models shows results inside
this experimental range.
In \cite{keis},  in a one-loop calculation, in the heavy baryon chiral 
perturbation theory, the author found $S=-2.8^o$,  in another work, \cite{meis}
using dispersion relations, the authors found $0^o\leq S\leq 1.1^o$, and in
\cite{ta1}, in the heavy baryon perturbation theory, the calculated phases were
 $-7.3^o\leq S\leq +0.4^o$,
$-3.5^o\leq S \leq +0.5^o$, that gives $-7.8^o\leq S-P\leq +3.9^o$
It would be very nice if the HyperCP data would determine a narrower 
experimental range of the phases.

\begin{acknowledgments}
I wish to tank M. R. Robilotta, for many helpful discussions.
This work was supported by FAPESP and CNPq.
\end{acknowledgments}

\appendix\section{Basic formalism}

In this paper $p$ and $p^{\prime}$ are the initial and final hyperon
4-momenta, $k$ and $k^{\prime}$ are the initial and final pion 4-momenta, so
the Mandelstam variables are 
\begin{eqnarray}
s &=& (p+k)^2=(p^{\prime}+k^{\prime})^2 \\
t &=& (p-p^{\prime})^2=(k-k^{\prime})^2 \\
u &=& (p^{\prime}-k)^2=(p-k^{\prime})^2 \ \ .
\end{eqnarray}
With these variables, we can define 
\begin{eqnarray}
\nu &=& {\frac{s-u}{4m}} \\
\nu_0 &=& {\frac{2\m^2-t}{4m}} \\
\nu_r &=& {\frac{m_r^2-m^2-k.k^{\prime}}{2m}} \ \ ,
\end{eqnarray}
where $m$, $m_r$ and $\m$ are, respectively, the hyperon mass, the 
resonance mass and the pion mass. The scattering amplitude for an isospin 
$I$ state is 
\beq
T=\overline{u}(\vec p\prime)\lbrace \lbrack A + 
{\frac{(\not\!k + \not\!k')}{2}}B\rbrack\rbrace u(\vec p)\ ,  
\eeq

\ni where $A$ and $B$ are calculated using the Feynman diagrams. So the 
scattering matrix is 
\beq
M^{ba} = {\frac{T^{ba}}{8\pi\sqrt{s}}} = f(\theta) + \vec\sigma.\hat n
g(\theta) = f_1 + {\frac{(\vec\sigma .\vec k' )(\vec\sigma .\vec k)}{kk'}}f_2
 \ \ , 
\eeq
with 
\begin{eqnarray}
& &f_1(\theta) = {\frac{(E+m)}{8\pi\sqrt{s}}} \lbrack A + (\sqrt{s}%
-m)B\rbrack\ \ , \\
& & f_2(\theta) = {\frac{(E-m)}{8\pi\sqrt{s}}} \lbrack -A + (\sqrt{s}%
+m)B\rbrack \ \ ,
\end{eqnarray}
where $E$ is the hyperon energy.

The partial-wave decomposition is done
with 
\beq
a_{l\pm} = {\frac{1}{2}}\int_{-1}^{1}\lbrack P_l(x)f_1(x) + P_{l\pm
1}(x)f_2(x)  \rbrack dx \ \ . 
\label{apar}
\eeq

In our calculation (tree level) $a_{l\pm}$ is real. With the unitarization, 
as explained in Section III, we obtain 
\beq
a_{l\pm}^U = {\frac{1}{2i}}\lbrack e^{2i\delta_{l\pm}} -1\rbrack =
e^{i\delta_{l\pm}}{\rm sen}(\delta_{l\pm})\rightarrow a_{l\pm} \ \ . 
\eeq

\section{loop integrals}

The basic loop integrals needed in order to perform the calculations of Fig. 2
are

\bea
&& I_{cc}^{\m\cdots} =  \int  \frac{d^4Q}{(2\pi)^4}\; 
\frac{\lp\frac{Q^\m}{\m}\cdots\rp}{[ (Q\!-\!q/2) ^2\! -\!\m^2] 
[(Q\!+\!q/2)^2 \!-\!\m^2]}
 \;,
\nonumber \\
&& \label{a`} \\
&& I_{\sb c}^{\m\cdots}= \int \frac{d^4Q}{(2\pi)^4}\;
\frac{\lp\frac{Q^\m}{\m}\cdots\rp}{[ (Q\!-\!q/2) ^2\! -\!\m^2] 
[(Q\!+\!q/2)^2 \!-\!\m^2]} \;
\frac{2m\m}{ [s -\mb^2 ] } \;. \nonumber \\
&&
\label{a2}
\eea

The integrals are dimensionless and have the following tensor structure
\bea
&& I_{cc} =  \frac{i}{(4\pi)^2} \lc \P_{cc}^{(000)}\rc \;,
\label{a3}\\
&& I_{cc}^{\m\n} = \frac{i}{(4\pi)^2}\lc \frac{q^\m q^\n}{\m^2} \; \P_{cc}^{(200)} 
+ g^{\m\n}\;\Pb_{cc}^{(000)}\rc \;,
\label{a4}\\
&& I_{\sb c} =  \frac{i}{(4\pi)^2}\lc \P_{\sb c}^{(000)} \rc\;,
\label{a5}\\
&& I_{\sb c}^{\m} =  \frac{i}{(4\pi)^2} \lc \frac{P^\m}{m} \; \P_{\sb c}^{(001)}\rc \;.
\label{a6}
\eea

Thus, the $\Pi$ integrals that appear in the text are
\bea
&& \P_{cc}^{(n00)} = - \int_0^1d a\;  (1/2-a)^n\; \ln \lp \frac{D_{cc}}{\m^2}\rp  \;,
\label{a7}\\[2mm]
&& \Pb_{cc}^{(000)}= - \;\frac{1}{2} \int_0^1 d a \; \frac{D_{cc}}{\m^2} \;
\ln \lp \frac{D_{cc}}{\m^2}\rp  \;,
\label{a8}\\[2mm]
&& \P_{\sb c}^{(00n)}=  \lp\!- 2m /\m \rp^{n+1} \int_0^1 d a\;a \int_0^1 d b \; 
\frac{\m^2\; (ab/2)^n}{D_{\sb c}}\;,  \nn \\
&&
\label{a9}
\eea

\ni
with
\bea
 D_{cc} &=& -a(1-a)\;q^2 + \m^2 \;,
\nn\\
 D_{\sb c} &=& -a(1-a)(1-b)\;q^2 \nn \\
&& + [\m^2 -ab\;(\m^2+m^2-\mb^2) + a^2b^2 \; m^2] \;.  
\nn
\eea


\end{document}